\documentstyle[11pt,epsf]{article}
\begin{document}

{\flushright\em Appeared in Science, 7 September 2001\\}

{\center\LARGE\sf Hydrogen 21-Centimeter Emission from\\ a Galaxy at 
Cosmological Distance\\}

{\center\large {\noindent
	M.~A.~Zwaan$^{1\star}$,
	P.~G. van Dokkum$^{2}$ \&
	M.~A.~W. Verheijen$^{34}$\\
	}}

\null\vspace{3mm}
\small{
	\raggedright$^{1}$ School of Physics, 
		University of Melbourne, Victoria 3010, Australia\\
  	\raggedright$^{2}$ California Institute of Technology,
		MS 105-24, Pasadena, CA 91125, USA\\
	\raggedright$^{3}$ Department of Astronomy, University of
		Wisconsin, 475 North Charter Street, \\
		\hspace{2.7mm}Madison, WI 53706, USA\\
	\raggedright$^{4}$ National Radio Astronomical Observatory, 
		PO Box 0, Socorro, NM 87801, USA\\
	\raggedright$^{\star}$ To whom correspondence should be
		addressed. E-mail: mzwaan@physics.unimelb.edu.au\\
	}
 \normalsize

\textwidth=13.7cm
\null\vspace{2mm}
{\bf 
 We have detected the neutral atomic hydrogen (HI) emission line at
a cosmologically significant distance ($z=0.18$) in the rich galaxy
cluster Abell 2218, with the Westerbork Synthesis Radio Telescope.  The
HI emission originates in a spiral galaxy $2.0\,h^{-1}_{65}$ Mpc from
the cluster core.  No other significant detections have been made in the
cluster, suggesting that the mechanisms that remove neutral gas from
cluster galaxies are efficient.  We infer that less than three gas-rich
galaxies were accreted by Abell 2218 over the past $10^9$ years.  This
low accretion rate is qualitatively consistent with low-density
cosmological models in which clusters are largely assembled at $z>1$. 
}

\textwidth=16.7cm
Galaxies in clusters have evolved in the past 2 -- 3 Gyr.  The number of
blue galaxies in clusters was higher in the past (the Butcher-Oemler
effect) ({\it 1, 2\/}), and spiral galaxies were more
prevalent ({\it 3-5\/}).  It has been argued that these
effects are caused by enhanced accretion of gas rich star forming
galaxies from the surrounding field ({\it 6-8\/}). 
Detailed modeling suggests that the neutral gas disks of in-falling
galaxies can be stripped by the hot x-ray gas that envelopes rich galaxy
clusters ({\it 9-11\/}).  Because the neutral gas provided
the fuel for star formation, the star formation rate drops precipitously
after the cold gas has been removed.  Hence galaxies rapidly fade and
redden after they have been accreted by a rich cluster.  The low neutral
atomic hydrogen (HI)
content of galaxies in the cores of the nearby Coma ({\it 12\/}) and
Virgo ({\it 13\/}) clusters is consistent with these models.  However,
at higher redshift, at which the galaxy accretion rate is predicted to be
higher and spiral galaxies are more abundant in the central regions of
rich clusters, these models have not been tested by direct observations
of the neutral gas reservoir of in-falling galaxies.  HI 21-cm emission
line studies have been limited to the local universe
({\it 12-15\/}) because radio synthesis telescopes were not
equipped to operate at frequencies corresponding to the redshifted HI
line, or lacked the sensitivity to detect the HI line at higher
redshifts. 

We have initiated a program of deep HI imaging of galaxy clusters
Abell 2218 and Abell
1689 at $z\sim 0.2$ to study the content and distribution of the 
HI in cluster galaxies at intermediate redshifts.  Here we report
on observations of Abell~2218 at $z=0.176$ from the recently upgraded
Westerbork Synthesis Radio Telescope (WSRT).  The cluster is extremely
rich and massive ({\it 16, 17\/}), has a luminous and
extended x-ray halo ({\it 18\/}), and has become widely known for the
Hubble Space Telescope imaging that revealed a rich structure of strong
gravitational arcs ({\it 19\/}). 

 Observations were performed with the WSRT during the commissioning of the
upgraded system in the period from July to September 1999.  Data were
taken with the cooled 21-cm receivers in two adjacent bands of 10~MHz
each, thus producing $2\times 128$ channels of each 78.1~kHz
corresponding to a velocity spacing of 19.5~km/s at the redshift of the
cluster and a resolution of 38.9~km/s after Hanning smoothing.  Each
frequency band was observed for 18$\times$12 hours, while the
positions of the four movable telescopes were varied. 
The results reported here are
based on the analysis of the usable 60\% of the data ({\it 20\/}). 
The data were taken around 1200~MHz, which is outside the protected
frequency bands for radio astronomy.  As a result, the data were
affected by human-made interfering signals, and careful inspection and
editing of the data was essential.  The spatial resolution in the final
data set is $18.0'' \times 19.7''$ and the r.m.s.  noise level after
Hanning smoothing is 0.11 mJy/beam in the lower frequency band and 0.10
mJy/beam in the higher frequency band.  The bandwidth and the primary
beam ($7\,h^{-1}_{65}$ Mpc diameter at FWHM) together define a total
survey volume of $\sim 2500\,h^{-3}_{65} \rm Mpc^3$. 

The most prominent signal in our data set amounts to $8\sigma$, with
optimal smoothing using a Gaussian filter with FWHM=20$''$ 
in the spatial domain and FWHM=60 km/s in the frequency domain (Fig.\ 1). 
No other significant ($>6 \sigma$) signals were found, neither in the 
full resolution nor the smoothed versions of the data cube.
 The integrated
flux in the detection, corrected for primary beam attenuation, is $33$
mJy km/s.  This is equivalent to an HI mass of $(5.4\pm 0.7) \times
10^9\, h^{-2}_{65}\,M_\odot$, which is less than the typical HI
mass of a field galaxy ($M_{\rm HI^*}=8.4 \times 10^9\,
h^{-2}_{65}\,M_\odot$) ({\it 21\/}).  The velocity width of the detected
emission line is small, $60 \pm 20$~km/s at 50\% of the peak
flux.  The narrowness of the signal explains why this modest HI mass
stands out from the noise.  The redshift of the HI line is
$z=0.1766$, coincident with the peak in the redshift distribution of the
confirmed cluster members ({\it 17\/}). 

To confirm the signal and to investigate the properties of the source, 
we used the Keck telescope to obtain optical imaging and spectroscopic
observations of the source responsible for the HI emission.  The optical
image with HI contours overlaid (Fig.  2) shows that the 
HI emission coincides with a spiral galaxy that has two well developed  
bluish spiral arms emanating from a redder and elongated NNE-SSW 
oriented bar-like structure. The western spiral arm runs along and
extends beyond a redder companion galaxy $\sim 18 h_{65}^{-1}$ kpc
to the southwest. 
We christen the HI selected spiral galaxy A2218-H1. 
The J2000 coordinates
of A2218-H1 are $\alpha$=16:33:58.5 and $\delta$= +66:10:06.
From the imaging observations we infer $R = 18.9 \pm 0.1$ mag 
for the spiral galaxy, which means that its intrinsic luminosity is 
about half that of the Milky Way Galaxy. For the 
companion galaxy we find $R=19.9 \pm 0.1$ mag.  Our
spectroscopic observations (Fig.  1) show that the optical redshift of A2218-H1
is $z=0.1766\pm 0.0001$ and that of the companion is $z=0.1768\pm
0.0002$, thus giving a velocity separation of $50\pm 60$~km/s.  Both
redshifts are within $1\sigma$ of the redshift of the HI detection. 
This confirms the identification and suggests that the spiral and its
companion are interacting. 
 Because of the limited spatial resolution of the HI measurements we can not
exclude the possibility 
that the companion galaxy contributes to the total measured HI
signal, although the small velocity width of the 
HI line suggests that the HI signal originates in a single galaxy.

The optical spectra indicate that A2218-H1 and its companion galaxy have
evolved stellar populations, and a low star formation rate although the
spiral arms of A2218-H1 are too faint to contribute much to its optical
spectrum.  The galaxies are not detected in our deep 1200 MHz continuum
map (rms noise $29\mu\rm Jy$), which provides a $3\sigma$ upper limit to
the star formation rate ({\it 22\/}) of $1.4 M_\odot\, \rm yr^{-1}$.  By
comparison, the current star formation rate of the Milky Way Galaxy is
about $4 M_\odot\, \rm yr^{-1}$ ({\it 23\/}).  Nearby interacting galaxies
in the field at $z=0$ generally show much higher star formation rates
({\it 24\/}), probably because the gas experiences shock-wave heating
during the interaction ({\it 25\/}).  Apparently star formation is
inhibited, even though all the conditions for a strong star burst seem
to be met: sufficient fuel and an interaction to trigger the burst. 

A2218-H1 does not show indications of the influence of interaction with
the intra-cluster medium (ICM).  The gas richness is typical of field
spiral galaxies ({\it 26\/}) ($M_{\rm HI}/L_R$=0.3), and within the
positional accuracy of 20 kpc, we detect no offset of the HI
distribution with respect to the stellar disk.  The most sophisticated
models of accretion onto clusters consist of three-dimensional smooth
particle hydrodynamics (SPH) simulations of spiral galaxies with a
complex multi-phase structure ({\it 11\/}).  These simulations predict
that the HI disk of A2218-H1 will be stripped completely in $1\times
10^8$ yr after the galaxy enters the ICM.  The position of A2218-H1 is
about 11 arcmin west from the central cD galaxy, which also marks the
peak of the x-ray profile.  This separation translates to a projected
distance of $2.0\,h^{-1}_{65}$ Mpc from the cluster core.  The detected
galaxy therefore resides in the outskirts of the cluster, beyond the
point where the bright x-ray halo ({\it 18\/}) has been measured.  The
small radial velocity difference between the cluster center and the
galaxy, in combination with the large projected distance suggests that
the galaxy is currently in-falling onto the cluster with a high radial
acceleration.  If the galaxy is on a trajectory towards the cluster
core, it will probably enter the ICM in $\sim 2\times 10^8$ yr
({\it 27\/}). 

Abell 2218 displays a moderate Butcher-Oemler effect: The blue galaxy
fraction in the core of the cluster is 11\% ({\it 1\/}).  None
of these blue galaxies have been detected in our 21-cm observations,
providing an average upper limit of $5.0 \times 10^9\,
h^{-2}_{65}\,M_\odot$ on the HI mass of individual galaxies
in the blue Butcher-Oemler
population, assuming a velocity width of 100 km/s.  
Although the brightest of the galaxies responsible for the
Butcher-Oemler effect have luminosities comparable to those of
normal field
spiral galaxies ({\it 1\/}), they must have lower gas-to-luminosity
ratios.  The low HI content of this extremely rich `Butcher-Oemler'
cluster provides support for current SPH models of the effects of ram
pressure stripping on the cold gas disks of galaxies.  In addition, 
the lack of a significant population of HI rich galaxies in the outskirts of
Abell 2218 implies that is has a low accretion rate of gas rich field
galaxies at the observed epoch.  Using the fact that the survey is
sensitive to galaxies with HI masses larger than $M_{\rm HI,*}$
throughout the primary beam, and the assumption that HI disks of
in-falling galaxies remain undepleted at distances from the cluster core
larger than that of A2218-H1, we can derive a 95\% confidence upper
limit to the accretion rate ({\it 28\/}) of $3\rm \,Gyr^{-1}$.  We
conclude that there is no large reservoir of gas rich galaxies that
might form a future `Butcher-Oemler' population, consistent with the low
Butcher-Oemler effect observed at $z=0$. 
The low accretion rate of this massive
cluster at $z \sim 0.2$ is in qualitative agreement
with low-density cosmological models in which clusters are largely
assembled at z>1 (e.g., {\it 6, 29\/}).

\newpage
\noindent {\bf References and Notes}
\begin{description}
\small

\item
 1. H. Butcher,  A. Oemler,
 {\it Astrophys.  J.} {\bf 285}, 426 (1984).

\item
 2.  W.~J. Couch, R.~M. Sharples,
 {\it Mon. Not. R.  Astron. Soc.} {\bf 229}, 423 (1987).

\item
 3. A. Dressler, 
 {\it Astrophys. J.} {\bf 490}, 577 (1997).

\item
 4.  W.~J. Couch,  A.~J. Barger, I. Smail, 
  R.~S. Ellis, R.~M. Sharples,
 {\it Astrophys.  J.} {\bf 497}, 188 (1998).

\item
 5. P.~G. van Dokkum, M. Franx, D. Fabricant,
  G.~D. Illingworth, D.~D.Kelson,
 {\it Astrophys. J.} {\bf 541}, 95 (2000).

\item
 6. G. Kauffmann,
  {\it Mon. Not. R.  Astron. Soc.} {\bf 274}, 153 (1995).

\item
 7. R.~G. Abraham et al.,
 {\it Astrophys. J.} {\bf 471}, 694 (1996).

\item
 8. B. Moore, N. Katz, G. Lake, A. Dressler, A. Oemler Jr.,
 {\it Nature} {\bf 379}, 613 (1996). 

\item
 9. M. G. Abadi, B. Moore, R.~G. Bower,.  
 {\it Mon. Not. R.  Astron. Soc.} {\bf 308}, 947 (1999).

\item
 10. M. Mori, A. Burkert, 
 {\it Astrophys.  J.} {\bf 538}, 559 (2000).

\item
 11. V. Quilis, B. Moore, R. Bower,  
 {\it Science} {\bf 288}, 1617 (2000).

\item
 12. H. Bravo-Alfaro,  V. Cayatte,  J.~H. van Gorkom,
 C. Balkowski,
 {\it Astron. J.} {\bf 119}, 580 (2000).

\item
 13. V. Cayatte, C. Balkowski, J.~H. van Gorkom,
  C. Kotanyi,
 {\it Astron.  J.} {\bf 100}, 604 (1990).

\item
 14. J.~M. Dickey,
 {\it Astron.  J.} {\bf 113}, 1939 (1997).

\item
 15. D.~G. Barnes, L. Staveley-Smith, R.~L. Webster,
 W. Walsh, 
 {\it Mon. Not. R.  Astron. Soc.} {\bf 288}, 307 (1997).

\item
 16. G.~O. Abell,  H.~G. Corwin Jr., R.~P. Olowin,
 {\it Astrophys. J. Suppl. Ser.} {\bf 70}, 1 (1989). 

\item
 17. J.~F. Le Borgne, P. Pell{\'o}, B. Sanahuja,
 {\it Astron. Astrophys. Suppl. Ser.} {\bf 95}, 87 (1992). 

\item
 18. G. Squires, et al.,
 {\it Astrophys.  J.} {\bf 461}, 572 (1996).

\item
 19. J.-P.Kneib,  R.~S. Ellis, I. Smail, 
  W.~J. Couch,  R.~M. Sharples,   
 {\it Astrophys.  J.} {\bf 471}, 643 (1996).

\item
 20.  Because these observations were
carried out during the commissioning of the upgraded WSRT, about
35\% of the data were by affected by software and hardware problems. 
Another 5\% of the data were affected by radio frequency
interference. 

\item
 21. M.~A. Zwaan, F.~H. Briggs, D. Sprayberry,
  E. Sorar,
 {\it Astrophys.  J.} {\bf 490}, 173 (1997).

\item
 22. L. Cram, M. Hopkins, B. Mobasher,  
 M. Rowan-Robinson,
 {\it Astrophys.  J.} {\bf 507}, 155 (1998).\\
 The relation given by these authors is valid for 1.4 GHz, whereas our
observations are at 1.2 GHz. 

\item
 23. X. Hern{\'a}ndez, V. Avila-Reese, C. Firmani,
 {astro-ph/} {\bf 0105092} (2001).

\item
 24. C.~T. Liu,  R.~C. Kennicutt Jr., 
 {\it Astrophys.  J.} {\bf 450}, 547 (1995).

\item
 25.  J.~C. Mihos, L. Hernquist,
 {\it Astrophys.  J.} {\bf 464}, 641 (1996).

\item
 26. M.~S. Roberts, M.~P. Haynes,
 {\it An. Rev. Astron. \& Astrophys.} {\bf 32}, 115 (1994).

\item
 27.  Here, we assume that A2218-H1 has an
in-fall velocity equal to the velocity dispersion of the cluster of 1370
$\rm km\, s^{-1}$ and has to travel $\sim$300 kpc to the edge of the
detectable x-ray halo. 

\item
 28.  The galaxy A2218-H1 shows no signs of
stripping so we can safely assume that the HI reservoirs of galaxies at
distances from the cluster center larger than that of A2218-H1, are not
depleted.  The annulus over which an in-falling $M_{\rm HI}>M_{\rm
HI,*}$ galaxy could be detected is therefore defined by the radius of
the primary beam and the radius of A2218-H1.  The width of this annulus
is 2 Mpc.  The maximum velocity of an infalling galaxy equals the
velocity dispersion of the cluster which is 1370 $\rm km\, s^{-1}$.  The
minimum time such a galaxy would be detectable is therefore $1.5\times
10^9$ yr.  Poisson statistics and the fact that only one galaxy is
detected give a 95\% upper limit to the number of galaxies within this
annulus of 4.5.  The 95\% upper limit to the infall rate is therefore
$3.1 \rm Gyr^{-1}$. 
This calculation is based on the assumption that A2218-H1 and
the cluster core are at the same distance from the observer.

\item
 29. E. Ellingson, H. Lin, H.~K.~C., Yee, 
  R.~G. Carlberg, 
  {\it Astrophys.  J.} {\bf 547}, 609 (2001).

\item
 30. J.~B. Oke, et al.,
 {\it Publ. Astron. Soc. Pac.} {\bf 107}, 375 (1995).

\item
 31. We thank the WSRT staff for assistance with the data taking and T.~Galama
and A.~Diercks for obtaining the Keck image.  A
significant part of the work of MAZ was carried out at the Kapteyn
Astronomical Institute, the Netherlands.  PGvD acknowledges support by
NASA through Hubble Fellowship grant HF-01126.01-99A awarded by the
Space Telescope Science Institute.  Until December 2000, MAWV was
employed by National Radio Astronomical Observatory 
through a Jansky Fellowship.  The WSRT is operated by
the Netherlands Foundation for Research in Astronomy (NFRA/ASTRON), with
financial support from the Netherlands Organization for Scientific
Research (NWO). 
\end{description}

\noindent 1 June 2001; accepted 27 July 2001

 \clearpage
 \normalsize

 \newpage
 \begin{figure}
 \epsfxsize=12.5cm \epsfbox[-60 210 380 680]{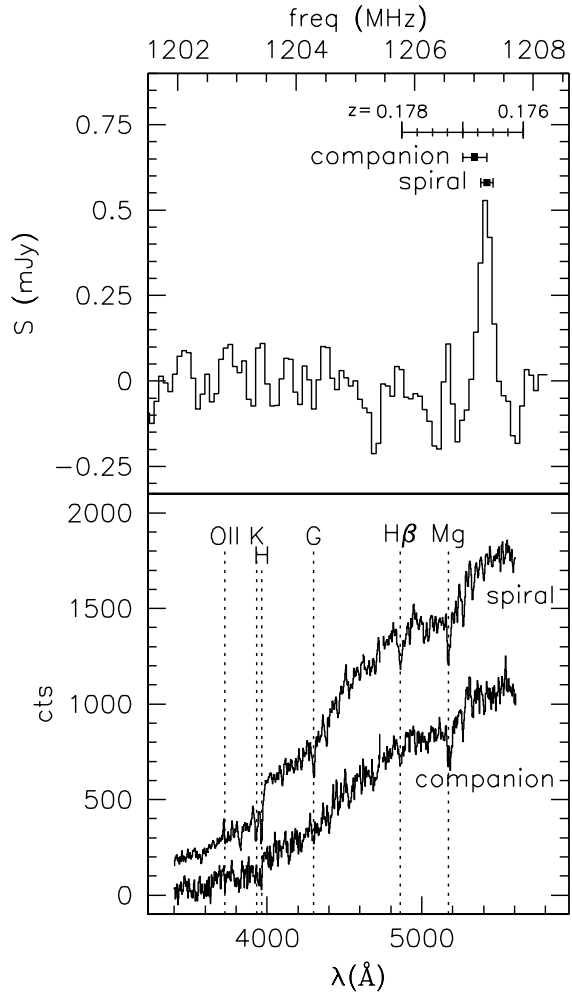}
 \caption{Spectra of the first HI selected galaxy at $z=0.18$. 
(A) The global HI profile.  It is Hanning smoothed which
results in a spectral resolution of 38.9 km/s.  The optical redshifts of
the spiral galaxy and its companion are also indicated, with $1\sigma$
uncertainties.  (B) The optical spectra of the spiral
galaxy A2218-H1 and its companion.  The spectra were obtained on 31 March 
2000 with the Low-Resolution Imaging Spectrometer ({\it 30\/}) on the
W.~M.  Keck I Telescope with the 300 lines $\rm mm^{-1}$ grating and a
$1''$ slit.  A2218-H1 was observed for 800 s, and its companion
for 300 s, both during twilight.  The redshift of A2218-H1 is
$z=0.1766\pm 0.0001$ and that of the companion galaxy 
is $z=0.1768\pm 0.0002$,
thus giving a velocity separation of $50\pm 60$~km/s.  Both redshifts
are within $1\sigma$ of the redshift of the HI detection. 
}
\end{figure}  

 \newpage
 \begin{figure}
 \epsfxsize=12cm \epsfbox[-160 210 250 580]{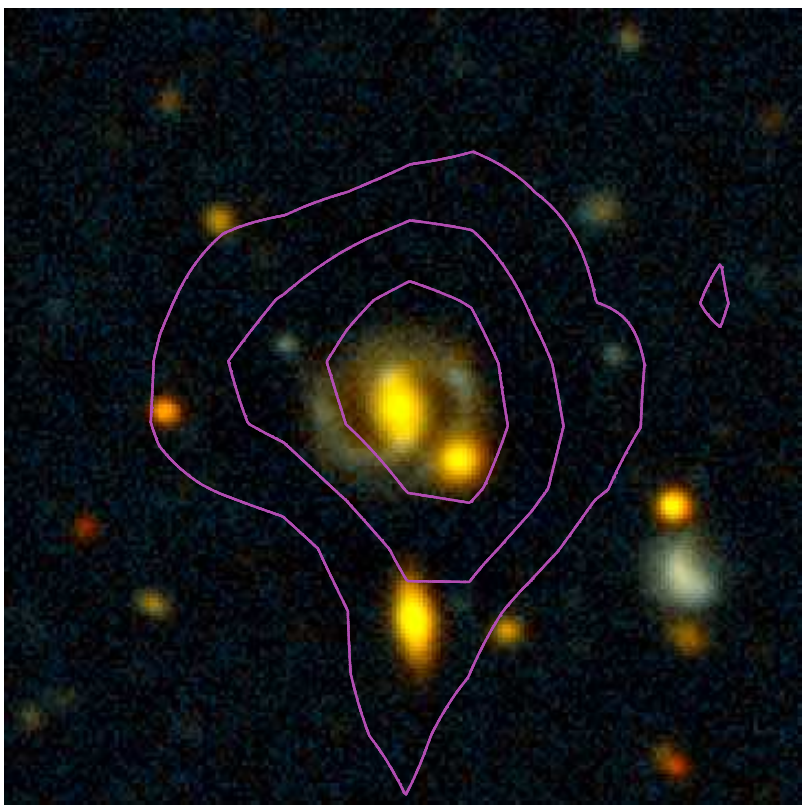}
 \caption{Overlay of HI contours on a color representation of the
optical image.  The contours correspond to $1.2$, $1.9$, and $2.5\times
10^{19}\,\rm cm^{-2}$.  In this image, the FWHM of the Westerbork
synthesized beam is $18.0'' \times 19.7''$, comparable to the extent 
of the second contour.
The optical imaging observations were
obtained on 4 April 2000 with the Echellette Spectrograph and Imager on
the W.~M.~Keck II Telescope.  The field was observed for 300 s in
the $R$-band, and 300 s in the $B$-band.  The seeing was $0.9''$. 
The size of the image is $60''\times 60''$ which corresponds to $184
\times 184\,h^{-1}_{65}$ kpc at the redshift of Abell 2218. }
\end{figure}

\end{document}